\documentclass[pre]{revtex4}
\usepackage{amsmath}
\usepackage{graphicx}
\begin{document}
\title{Widening Disparity and its Suppression in a Stochastic Replicator Model}
\author{Hidetsugu Sakaguchi}
\affiliation{Department of Applied Science for Electronics and Materials,
Interdisciplinary Graduate School of Engineering Sciences, Kyushu
University, Kasuga, Fukuoka 816-8580, Japan}
\begin{abstract}
Winner-take-all phenomena are observed in various competitive systems. We find similar phenomena in replicator models with randomly fluctuating growth rates. The disparity between winners and losers increases indefinitely, even if all elements are statistically equivalent. A lognormal distribution describes well the nonstationary time evolution. If a nonlinear load corresponding to progressive taxation is introduced, a stationary distribution is obtained and disparity widening is suppressed.
\end{abstract}
\maketitle
\section{Introduction}
One topic in econophysics is the statistical distributions of wealth and income. 
The Pareto-like power law distributions have been studied using mathematical models~\cite{Bouchaud,Yakovenko}. 
In the power law distribution of wealth, a small percentage of wealthy people possess a large amount of money~\cite{Piketty}. 
Winner-take-all phenomena are closely related phenomena in which a few winners occupy almost all the market share in a modern economy~\cite{Arthur,Frank}. Very small differences in performance lead to large differences in market share. One classical example is that ``QWERTY" has become the standard layout of computer keyboards. 

The replicator model is a model that can describe the competition and selection in evolutionary game dynamics~\cite{Taylor,Hofbauer}. The replicator equation expresses the dynamics of the population $x_i$ for the strategy $i$ with the fitness $f_i$. In the simplest model, only the fittest species wins the competition and occupies the whole population. It can be applied to the molecular evolution of quasispecies~\cite{Eigen,Nowak}. 
In the usual replicator model, the fitness is randomly distributed but constant in time. Several authors studied stochastic replicator models to discuss the effect of fluctuations on the Nash equilibrium in game theory~\cite{Fundenberg,Foster,Imhof}. Closely related stochastic Lotka-Vorterra models were also studied by several authors~\cite{Kirlinger}. For example, Mao et al. discussed the suppression of population explosion by environmental noises~\cite{Mao}. 

In this paper, we study a very simple replicator model with a randomly fluctuating growth rate as a model of widening disparity. The origin of widening disparity in a modern economy was discussed by several authors using some positive feedback effects such as the law of ``$r>g$" (the return on capital is larger than the income growth rate)~\cite{Piketty} or the law of increasing returns~\cite{Arthur}. We show that winner-take-all phenomena occur in the simplest model, and that the disparity among many elements increases indefinitely even if each element is statistically equivalent. Then, we introduce a nonlinear load factor corresponding to progressive taxation and show that disparity widening is suppressed. Furthermore, we find stationary equilibrium distributions for the Fokker-Planck equation. 

\section{Stochastic Replicator Model and Winner-Take-All Phenomena}
We consider a stochastic replicator model with fluctuating growth rates. 
The model equation for $N$ replicators is written as
\begin{equation}
\frac{dx_i}{dt}=(r_i(t)-c)x_i(t),\;\;\; {\rm for}\;i=1,2,\cdots,N,
\end{equation}
where $r_i(t)$ is assumed to be a Gaussian white noise satisfying $\langle r_i(t)\rangle=0$ and $\langle r_i(t)r_j(t^{\prime})\rangle=2T\delta_{i,j}\delta(t-t^{\prime})$, and $c$ is the average of $r_i$ expressed as
\begin{equation}
c=\frac{\sum_{i=1}^Nr_i(t)x_i(t)}{\sum_{i=1}^Nx_i(t)}.
\end{equation}
The sum $X=\sum_{i=1}^Nx_i$ is maintained to be constant in this model. The competition occurs owing to the term including $c$. If $x_i(0)$'s are all positive, $x_i(t)>0$ is satisfied for any $t>0$ in Eq.~(1). 
We interpret the coupled stochastic differential equation in the Stratonovich sense. Under the interpretation in the Stratonovich sense, stochastic variables can be treated as ordinary smooth variables, therefore, the Stratonovich stochastic equation is often used in physics. In mathematics and economics, the stochastic differential equation in the Ito sense is used more frequently. Related stochastic difference equations are sometimes used in econophysics, in which no ambiguity of interpretation appears.~\cite{Sato}   

First, we show a numerical result at $N=10000,T=0.005$. The initial condition $x_i(0)$ is a random number between 0 and 1. The total sum $X$ is fixed to be $N/2$. We have performed numerical simulation of Eq.~(1) by the Heun method~\cite{ST}.  
Figure 1 shows profiles of $x_i$ at $t=100,200,\cdots1000$.
Each element is statistically equivalent; however, winners and losers appear and the disparity increases with time. Lucky elements, whose $r_i$'s take large values for a long time, take almost all. Figure 2(a) shows the time evolution of the disparity $G$ (the Gini coefficient) defined by 
\begin{equation}  
G=\frac{\sum_{i=1}^N\sum_{j=1}^N|x_i-x_j|}{2\mu N^2},
\end{equation}
where $\mu=X/N$ is the average of $x_i$. 
Figure 2(b) shows the time evolution of the occupancy rate $R$ of the highest group defined by 
\begin{equation}
R=\frac{\sum_{i}x_i}{X},
\end{equation}
where the summation is taken from the largest to the 100th largest $x_i$. 
$R$ is the occupancy rate of the top one percent. $R$ increases with time and approaches 1. These results suggest that a winner-take-all phenomenon occurs and the disparity increases indefinitely in our model. This stochastic process seems to be 
nonstationary. 
\begin{figure}
\begin{center}
\includegraphics[height=4.cm]{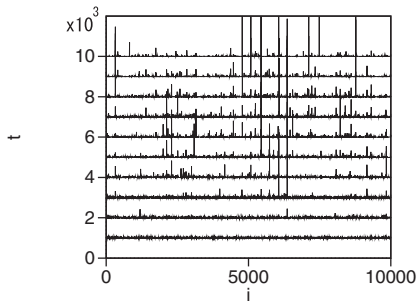}
\end{center}
\caption{Profiles of $x_i$ at $t=100,200,\cdots1000$. }
\label{f1}
\end{figure}
\begin{figure}
\begin{center}
\includegraphics[height=4.cm]{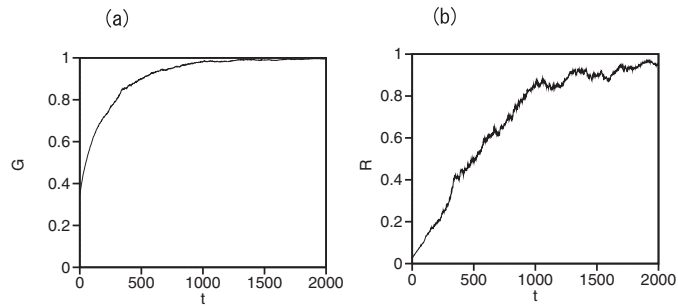}
\end{center}
\caption{(a) Time evolutions of the disparity $G$. (b) Time evolution of the occupancy rate $R$ of the top one percent group.}
\label{f2}
\end{figure}

To understand the nonstationary time evolution, we first consider a simpler model of $N=2$. For $N=2$, Eq.~(1) becomes
\begin{equation}
\frac{dx_1}{dt}=\left [ r_1(t)-\frac{x_1r_1(t)+x_2r_2(t)}{x_1+x_2}\right ] x_1.
\end{equation}
The ratio $q=x_1/(x_1+x_2)$ obeys
\begin{equation}
\frac{dq}{dt}=\{r_1(t)-r_2(t)\}q(1-q).
\end{equation}
This equation is a kind of stochastic logistic equation. 
Equation (6) is solved as
\begin{equation}
\frac{q(t)}{1-q(t)}=\frac{q(0)}{1-q(0)}\exp\left\{\int_0^t(r_1(t)-r_2(t))dt\right\}.
\end{equation}
If $q(0)=1/2$ is assumed for simplicity, the probability distribution of $\int_0^t(r_1(t)-r_2(t))dt$ obeys the Gaussian distribution of variance $4Tt$. The probability of $q(t)$ is given by 
\begin{equation}
P(q)=\frac{1}{q(1-q)}\frac{1}{\sqrt{8\pi Tt}}e^{-[\log\{q/(1-q)\}]^2/(8Tt)}.
\end{equation}
Figure 3 shows the probability distribution at $t=10,1000$, and 100000 for $T=0.005$. $P(1/2)$ decays as $1/\sqrt{t}$, and $P(0+)$ and $P(1-)$ increase indefinitely for $t\rightarrow \infty$. That is, one of two elements tends to occupy the whole.  

On the other hand, for $N=\infty$, $y_i(t)=\log(x_i(t))$ obeys
\begin{equation}
\frac{dy_i}{dt}=r_i(t)-c(t).
\end{equation}
The probability distribution of $y_i(t)$ is given by
\begin{equation}
P(y_i)=\frac{1}{\sqrt{4\pi Tt}}\exp\left\{-\left (y_i-y(0)+\int_0^t c(t^{\prime})dt^{\prime}\right )^2/(4Tt)\right \}.
\end{equation}
If $x_i(0)=\mu=X/N$ is assumed for all $i$, $y_i(0)=\log\mu$. The parameter $c$ is given by the condition that the average value $\langle x(t)\rangle$ is constant in time and takes the value $\mu$. The average value is calculated as 
\begin{eqnarray}
\langle x\rangle&=&\int_0^{\infty} xP(x)dx=\int_{-\infty}^{\infty}e^yP(y)dy\nonumber\\
&=&\int_{-\infty}^{\infty}\frac{1}{\sqrt{4\pi Tt}}\exp(y)\exp\left \{-\left (y_i-\log\mu+\int_0^t c(t^{\prime})dt^{\prime}\right )^2/(4Tt)\right \}dy\nonumber\\
&=&\mu\exp\left \{Tt-\int_0^tc(t^{\prime})dt^{\prime}\right \}.
\end{eqnarray}
The parameter $c$ needs to satisfy $c=T$, because $\langle x\rangle=\mu$ in Eq.~(11). Because the temporal average of growth rate $r_i(t)-c$ is $-T$, $x_i$'s of almost all elements decay to zero. 
The probability distribution $P(x,t)$ is expressed as
\begin{equation}
P(x,t)=\frac{1}{x}\frac{1}{\sqrt{4\pi Tt}}\exp\left \{-\left (\log x-\log\mu+Tt\right )^2/(4Tt)\right \}.
\end{equation}
This is a lognormal distribution. The average of $y$ decreases with $\log\mu-Tt$ and the variance of $y$ increases as $4Tt$. The average of $x$ is fixed to be constant but the average of $\log x$ decreases in proportion to $t$, which  might be a simple mathematical description of a winner-take-all phenomenon. 

Figure 4(a) shows $P(x)$ at $t=20,100$, and 2500 in a double-logarithmic plot for $T=0.005$ and $\mu=0.5$. At $t=20$, $P(x)$ takes a one-hump structure. As $t$ increases, $P(x)$ tends to take a power law distribution. Figure 4(b) shows a probability larger than $\mu/2=0.25$ as a function of $t$. The dashed line is the probability calculated from Eq.~(12) and the solid line denotes numerical results of Eq.~(1) with $N=10000$. Good agreement is observed. 
The probability that $x$ is larger than half of the average decreases rapidly, because winners take almost all. 

\begin{figure}
\begin{center}
\includegraphics[height=4.cm]{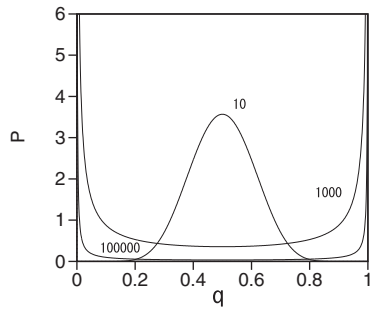}
\end{center}
\caption{Probability distributions expressed by Eq.~(8) at $t=10,1000$, and 100000 for $T=0.005$.}
\label{f3}
\end{figure}
\begin{figure}
\begin{center}
\includegraphics[height=4.cm]{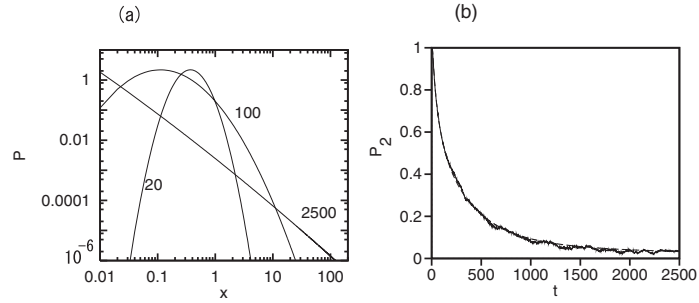}
\end{center}
\caption{(a) Probability distributions expressed by Eq.~(12) at $t=20,100$, and 2500 for $T=0.005$ and $\mu=0.5$. (b) Probability $P_2=\int_{\mu/2}^{\infty} P(x)dx$ for $T=0.005$ and $\mu=0.5$ obtained by direct numerical simulation of $N=10000$ (solid line) and using Eq.~(12) (dashed line).}
\label{f4}
\end{figure}
\section{Suppression of Disparity by Nonlinear Loads}
One of the effective methods of the income redistribution is progressive taxation~\cite{Piketty}. We can introduce a load term corresponding to progressive taxation. Equation (1) is replaced with the following model equation: 
\begin{equation}
\frac{dx_i}{dt}=\{r_i(t)-\alpha x_i^{\beta}-c\}x_i(t),\;\;\; {\rm for}\;i=1,2,\cdots,N,
\end{equation}
where $\alpha>0$, $\beta>0$, and $c=\{\sum_{i=1}^N(r_i(t)-\alpha x_i^{\beta})x_i(t)\}/X$. Here, the load in proportion to $x_i^{\beta+1}$ is assumed as a simple model.     
The case $\beta=0$ corresponds to tax in proportion to income. In this case, disparity widening cannot be suppressed, because the constant factor $\alpha$ is renormalized into $c$, and no effect appears in our model.  The case $\beta>0$ corresponds to progressive taxation. Loads heavier than that in proportion to $x_i$ are imposed on richer elements.  

The variable $y_i=\log x_i$ obeys
\begin{equation}
\frac{dy_i}{dt}=r_i(t)-\alpha e^{\beta y_i} -c,\;\;\; {\rm for}\;i=1,2,\cdots,N,\end{equation}
This is the Langevin equation in the potential $U=\alpha e^{\beta y}/\beta+cy$. The stationary distribution of the Fokker-Planck equation for $P(y)$ is given by
$P(y)\propto e^{-U/T}$. From $P(y)$, the stationary distribution of $x$ is calculated as 
\begin{equation}
P(x)\propto x^{-1-c/T}e^{-\alpha x^{\beta}/(\beta T)}.
\end{equation}
This stationary distribution can also be obtained directly as a stationary solution of the Fokker-Planck equation for $P(x,t)$: 
\[\frac{\partial P}{\partial t}=\frac{\partial}{\partial x}\left [(\alpha x^{\beta+1}+cx)P+Tx\frac{\partial}{\partial x}(xP)\right ].\]
\begin{figure}
\begin{center}
\includegraphics[height=4.cm]{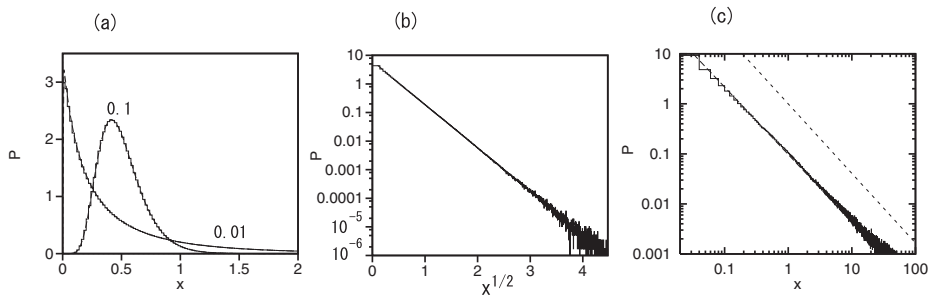}
\end{center}
\caption{(a) Stationary distributions $P(x)$ at $\alpha=0.1$ and 0.01 for $T=0.005,\langle x\rangle=0.5$, and $\beta=0.5$. (b) Semilogarithmic plot of $P(x)$ at $\alpha=\sqrt{12}(T/2)$ for $T=0.005,\langle x\rangle=0.5$, and $\beta=0.5$. The horizontal axis is $x^{1/2}$. (c) Double-logarithmic plot of $P(x)$ at $\alpha=0.001$, $T=0.005,\langle x\rangle=0.5$, and $\beta=0.05$.}
\label{f5}
\end{figure}
The average $\langle x\rangle=X/N$ is given by
\begin{equation}
\langle x\rangle=\frac{\int_0^{\infty}x^{-c/T}e^{-\alpha x^{\beta}/(\beta T)}dx}{\int_0^{\infty}x^{-1-c/T}e^{-\alpha x^{\beta}/(\beta T)}dx}=\left (\frac{\beta T}{\alpha}\right )^{1/\beta}\frac{\Gamma(1/\beta-c/(\beta T))}{\Gamma(-c/(\beta T))}.
\end{equation}
The parameter $c$ is determined from Eq.~(16). For $\beta=1/2$, Eq.~(16) is reduced to 
\begin{equation}
\langle x\rangle=\left (\frac{T}{2\alpha}\right ) \left (1-\frac{2c}{T}\right )\left (-\frac{2c}{T}\right ).
\end{equation}
Figure 5(a) shows the stationary distributions at $\alpha=0.1$ and 0.01 for $\beta=0.5,\langle x\rangle=0.5$, and $T=0.005$. The histograms show numerical results and the dashed lines denote Eq.~(15). Good agreement is seen. At $\alpha=0.1$, the distribution is localized at around $x=0.5$. 
This implies that the income is redistributed by progressive taxation, and disparity widening is suppressed. 
At $\alpha=0.01$, the distribution has a long tail.  
If $c<-T$, $P(x)=0$ at $x=0$, and a single peak appears in $P(x)$. 
If $-T<c<0$, $P(x)$ increases infinitely like $1/x^{1+c/T}$ as $x$ approaches 0. For $\beta=1/2$, $c=-T$ is realized at $\alpha=\alpha_c=(T/2)\sqrt{6/\langle x\rangle}$.
At $\alpha=\alpha_c$, $P(x)$ decays as $P(x)\propto e^{-\alpha x^{\beta}/(\beta T)}$, which is the stretched exponential distribution. Figure 5(b) shows the stationary distribution at $\alpha=\alpha_c=0.00866$ for $T=0.005, \beta=0.5$, and $\langle x\rangle=0.5$. The horizontal axis is $x^{1/2}$ and the vertical axis is $\log_{10} P$. The straight line implies the stretched exponential distribution. When both $\beta$ and $\alpha$ are small,  $e^{-\alpha x^{\beta}/(\beta T)}$ is almost constant at a small $x$ and the power law distribution $P\sim 1/x^{1+c/T}$ dominates. Figure 4(c) shows the stationary distribution in a double-logarithmic plot at $\beta=0.05,\alpha=0.001,\langle x\rangle=0.5$, and $T=0.005$. 
The dotted line denotes a power law of the exponent $1.39$. 
The stationary distribution is approximated at the power law distribution.
In the limit of $\alpha=0$, $c=0$ and $P(x)\sim 1/x$. However, the distribution is not stationary, because the normalization is impossible.  
In numerical simulation, a winner-take-all phenomenon occurs, and the disparity coefficient increases to 1 for $t\rightarrow\infty$, as shown in the previous section.  
\section{Summary}
We have proposed a stochastic replicator model as a model of the competing zero-sum world. A winner-take-all phenomenon is observed in direct numerical simulation. The Geni coefficient and occupancy rate increase to 1 with time. 
Widening disparity occurs naturally even if each element is statistically equivalent and no positive feedback effect such as the law of increasing returns~\cite{Arthur} is assumed. For $N=\infty$, the time evolution of $x$ is shown to obey the lognormal distribution in which the average value of $y=\log x$ decreases with $-Tt$. The probability distribution is not stationary. The average of $x$ is fixed to be constant but the average of $\log x$ decreases in proportion to $t$, which might be a simple mathematical expression of a winner-take-all phenomenon. 

When a nonlinear load factor is introduced, disparity widening is suppressed and a stationary distribution of $x$ is obtained. When the parameters $\alpha$ and $\beta$ are large, the distributions have a peak structure around the average. When they are small, the stationary distributions have a long tail for large $x$. When $\alpha$ is smaller than the critical value $\alpha_c$, the peak structure disappears and $P(x)$ increases in accordance with a power law as $x$ approaches 0. At $\alpha=\alpha_c$, a stretched exponential distribution is obtained. 

Our model is a coupled system of $N$ elements through the term including $c$. For $N=\infty$, our model becomes a kind of mean-field model. Because of the simplicity of our model, the probability distributions could be explicitly obtained. However, more realistic models including complicated terms would be necessary for the application to real economy, which is left for future study.  

\end{document}